\documentclass[12pt,tightenlines,eqsecnum,floats,showpacs,nofootinbib,amsmath,amssymb,aps,prd,superscriptaddress]{revtex4-1}

\usepackage{graphicx}
\usepackage{amsmath,amssymb}
\usepackage{hyperref}

\begin{document}
	
\title{Unitary evolution and uniqueness of the Fock representation of Dirac fields in cosmological spacetimes}
	
\author{Jer\'onimo Cortez}
\email{jacq@ciencias.unam.mx}
\affiliation{Departamento de F\'isica, Facultad de Ciencias, Universidad Nacional Aut\'onoma de M\'exico, M\'exico D.F. 04510, M\'exico}
\author{Beatriz Elizaga Navascu\'es}
\email{beatriz.elizaga@iem.cfmac.csic.es}
\affiliation{Instituto de Estructura de la Materia, IEM-CSIC, Serrano 121, 28006 Madrid, Spain}
\author{Mercedes Mart\'in-Benito}
\email{m.martin@hef.ru.nl}
\affiliation{Radboud University Nijmegen, Institute for Mathematics, Astrophysics and Particle Physics, Heyendaalseweg 135, NL-6525 AJ Nijmegen, The Netherlands}
\author{Guillermo A. Mena Marug\'an} \email{mena@iem.cfmac.csic.es}
\affiliation{Instituto de Estructura de la Materia, IEM-CSIC, Serrano 121, 28006 Madrid, Spain}
\author{Jos\'e M. Velhinho}
\email{jvelhi@ubi.pt}
\affiliation{Dept. de F\'isica, Universidade da Beira Interior, 6201-001 Covilh\~a, Portugal}

\begin{abstract}

We present a privileged Fock quantization of a massive Dirac field in a closed Friedmann-Robertson-Walker cosmology, partially selected by the criteria of invariance of the vacuum under the symmetries of the field equations, and unitary implementation of the dynamics. When quantizing free scalar fields in homogeneous and isotropic spacetimes with compact spatial sections, these criteria have been shown to pick out a unique Fock representation (up to unitary equivalence). Here, we employ the same criteria for fermion fields and explore whether that uniqueness result can be extended to the case of the Fock quantization of fermions. For the massive Dirac field, we start by introducing a specific choice of the complex structure that determines the Fock representation. Such structure is invariant under the symmetries of the equations of motion. We then prove that the corresponding representation of the canonical anticommutation relations admits a unitary implementation of the dynamics. Moreover, we construct a rather general class of representations that satisfy the above criteria, and we demonstrate that they are all unitarily equivalent to our previous choice. The complex structures in this class are restricted only by certain conditions on their asymptotic behavior for modes in the ultraviolet sector of the Dirac operator. We finally show that, if one assumes that these asymptotic conditions are in fact trivial once our criteria are fulfilled, then the time-dependent scaling in the definition of the fermionic annihilation and creation-like variables is essentially unique.
\end{abstract}

	
\maketitle

\section{Introduction}
\label{sec:Intro}

In general, the process of quantizing a classical system is plagued with ambiguities that often lead to different quantum theories, with different physical predictions. These ambiguities in the quantum representation persist even if one selects a specific Poisson algebra of classical observables and demands it to be irreducibly represented in the quantum theory. In the case of physical systems with a finite number of degrees of freedom, a powerful result comes at hand to remove this ambiguity. This is the Stone-von Neumann uniqueness theorem \cite{svn}, which states that all strongly continuous, unitary, and irreducible representations of the Weyl algebra are unitarily equivalent, and hence provide the same physics. However, this result does not apply in the case of field-like systems, and a priori there exists an infinite number of unitarily inequivalent representations of the field analogue of the Weyl algebra. 

This situation persists even if one restricts the attention to free field theories and Fock quantizations of them \cite{wald}. There, the ambiguity resides in the choice of the one-particle Hilbert space, or equivalently, in the choice of vacuum. Specifically, one defines annihilation and creation-like variables, in terms of which the solutions of the field equations are expressed. Their representation as annihilation and creation operators fully characterizes the quantum theory. The physically relevant freedom available in this choice of variables is encoded in the so-called \emph{complex structure} \cite{wald}, a real linear map on the space of solutions whose square is minus the identity. In the case of a scalar field, the complex structure has to preserve the symplectic form and, combined with it, must provide an inner product on phase space. The ambiguity in the quantization may be placed in the selection of the inner product from which the one-particle Hilbert space of the theory is constructed. On the other hand, when the field is a Dirac spinor, one may argue that there is no such an ambiguity in the inner product, inasmuch as there exists a natural choice in the space of solutions of the Dirac equations \cite{Dimock}. The complex structure needs to be compatible with this inner product. Here the ambiguity lies in the infinitely many possible splittings of the (complexification of the) space of solutions into particle and antiparticle subspaces \cite{bratteli}. In both the scalar and the Dirac field cases, different choices of complex structures lead to different sets of creation and annihilation-like variables. These different sets of variables are related by linear canonical transformations, usually called Bogoliubov transformations. The physical ambiguity is reflected in the fact that not all of these Bogoliubov transformations can be implemented unitarily in the quantum theory, which in turn is equivalent to say that the Fock representations that they relate are not unitarily equivalent.

Despite these problems, one often may impose some physical criteria in order to select a privileged class of Fock quantizations \cite{wald,ashmag,flor}. For free (test) fields propagating in globally hyperbolic spacetimes, the usual strategy is to demand a unitary implementation of the group of isometries of the considered background, or alternatively of the symmetries of the field equations on that background. The most natural way to obtain this is by imposing invariance of the complex structure, and hence of the vacuum, under the action of those symmetries. In certain scenarios, such as in stationary spacetimes, this restriction suffices to pick out a unique equivalence class of Fock representations. For instance, this is the case in Minkowski spacetime, where a unique vacuum is selected by demanding Poincar\'e invariance \cite{wald, parker}. However, in more generic scenarios, such as for cosmological spacetimes, stationarity is absent, and imposing just the invariance of the vacuum under the remaining symmetries of the system is generally not enough to fix the ambiguities in the Fock quantization. At this stage, a natural way to proceed is to replace the demand of invariance under time evolution with the requirement that the dynamics be implementable in terms of a unitary operator in the quantum theory. From a physical point of view, this condition restricts the allowed representations to those in which the vacuum, even if evolving in time, at least undergoes just a finite creation of particles during any finite period of the evolution. In this sense, unitarity of the evolution actually imposes ultraviolet regularity conditions \cite{AOP}.

These criteria of symmetry invariance of the vacuum and unitary implementation of the dynamics have been applied recently to the quantization of scalar fields, propagating in several types of cosmological spacetimes with compact spatial sections, leading to the remarkable result of ensuring uniqueness in the Fock quantization. Indeed, the two requirements suffice to select a unique canonical pair of field variables among all those that are related by homogeneous time-dependent scalings, as well as a unique (up to unitary equivalence) Fock representation for the canonical commutation relations of this pair. Such uniqueness theorem was first proven for the Fock quantization of scalar fields in Gowdy cosmologies  \cite{gowdy}. These are dimensional reductions of General Relativity with two spatial Killing vectors and compact spatial sections \cite{gowdyorig}. After a partial gauge fixing, the system reduces to that of a scalar field propagating in a two dimensional spacetime. Afterwards, the uniqueness of the quantization was also demonstrated for test scalar fields propagating in homogeneous and isotropic expanding backgrounds with spatial sections isomorphic to $d$-spheres, with $d<4$ \cite{spheres}. More recently, this latter result has been generalized to all possible compact topologies, as long as the spatial dimension is smaller than or equal to three \cite{compact}. In particular, this includes the case of a three-torus topology \cite{torus}, that is of special interest in modern cosmology, because it is the one of a flat universe, the case most favored by actual observations \cite{cmb}.

These results constitute an important improvement in the unambiguous construction of quantizations of scalar fields in curved spacetimes, especially when these are homogeneous and isotropic cosmologies. This is of crucial relevance when it comes to analyze the possible quantum phenomena in the primordial epochs of the universe. In fact, nowadays there is the opportunity to test physical predictions of the theoretical models by comparison with precise measurements that are being made in the best studied observational window to those early times: The cosmic microwave background \cite{cmb}. Nevertheless, and despite the interest that scalar types of matter may have in these situations (e.g. in the study of primordial perturbations), the physical phenomena that take place in the framework of the Standard Model of particle physics involve other types of fields, such as gauge fields or fermion fields. It is therefore natural to explore the behavior of these, in some sense more realistic, types of quantum fields in the context of the early universe. These questions have already been addressed in other works during the last decades. In particular, Ref. \cite{H-D} analyzes a specific Fock quantization for fermionic perturbations over a quantum inflationary closed Friedmann-Robertson-Walker (FRW) cosmology. From the point of view commented above about the possible choices of Fock representation, the discussion presented in Ref. \cite{H-D} contains an interesting result: During inflation, there is a finite creation of fermionic particles from the vacuum. This finiteness can actually be traced back to the fact that the selected Fock representation admits a unitary implementation of the dynamics, as we will show in this work.

A consistent answer to the question of whether the quantum effects of fermionic matter may be relevant in cosmology, or in more general situations, necessarily requires a robust description of the Fock quantization of a Dirac spinor in such curved spacetimes. With this motivation, here we present an analysis of the Fock quantization of a fermion field in the cosmological scenario analyzed in Ref. \cite{H-D}. More specifically, and keeping in mind the criteria employed for the case of scalar fields in Refs. \cite{gowdy,spheres,compact,torus}, we will demonstrate that the complex structure selected in Ref. \cite{H-D} not only is invariant under the symmetries of the dynamical field equations (which include the isometries of the spatial sections, that have the topology of the three-sphere $S^{3}$), but also admits a unitary implementation of the evolution, as we already pointed out. Since it does not mix modes that are dynamically decoupled, this complex structure turns out to admit a block-diagonal matrix form in a basis of eigenmodes of the Dirac operator on $S^{3}$.  Furthermore, it has a rather specific asymptotic ultraviolet behavior with respect to the eigenvalues of that operator. Fixing the usual convention of assigning the concepts of particle and antiparticle to, respectively, the eigenspace of the complex structure with eigenvalue $+i$ and the complex conjugate of the eigenspace with eigenvalue $-i$, we will also prove that any other complex structure that is selected by our physical criteria of invariance and unitarity, and that possesses the same kind of ultraviolet asymptotics as the one in Ref. \cite{H-D}, must be related to the latter by a unitarily implementable transformation. Hence our results show that all Fock representations allowed by our criteria, and that admit a structure with the specified ultraviolet asymptotics, form an equivalence class under unitary transformations. Within this class, we pick out the simplest one to serve as a {\it reference} quantization. In consequence, our reference vacuum (or, e.g., the vacuum chosen in Ref. \cite{H-D}) for the quantization of the Dirac field in a closed FRW cosmology is the only one, up to unitary equivalence, that is consistent with invariance under the symmetries of the dynamical field equations and with the requirement of a unitary evolution, if one restricts all considerations to the mentioned class of complex structures. This is a quite general, but not yet complete uniqueness result for the Fock quantization, inasmuch as we do not prove here that the considered class covers in fact all possible complex structures, namely, that there cannot exist any complex structure satisfying our criteria which nevertheless displays a different ultraviolet asymptotic dependence. However, if such a general result about uniqueness is valid, we argue in this work that, up to irrelevant terms, the time-dependent scaling in the definition of annihilation and creation-like variables of our reference quantization is the unique one that is allowed in the whole privileged equivalence class of quantizations that our criteria determine. Interestingly, and displaying a key difference with the scalar field case studied in Refs. \cite{gowdy,spheres,compact,torus}, this time-dependent scaling is not a global factor affecting the whole fermion field, but scales differently its particle and antiparticle parts.

The paper is organized as follows. In Sec. \ref{sec:Model} we revisit the classical cosmological model with a fermion field investigated in Ref. \cite{H-D}, and we propose a similar but simpler complex structure for its Fock quantization, which will serve us as a reference one in our discussion. In Sec. \ref{sec:Unitarity} we determine which complex structures are allowed under the requirement of invariance under the symmetries of the field equations. We will call those structures {\it invariant}. Besides, we study the dynamical evolution, seen as a Bogoliubov transformation, of the annihilation and creation-like variables defined by the invariant complex structures, and deduce the restrictions that the unitarity of the dynamics imposes on them. We then check that both our reference quantization and the one chosen in Ref. \cite{H-D} satisfy these requirements of invariance and unitary evolution. In Sec. \ref{sec:uniqueness}, we prove that any other complex structure that fulfills those requirements and displays the same ultraviolet asymptotic behavior (as ours or as the one in Ref. \cite{H-D}) is related to our reference complex structure by a unitary transformation. In addition, and beyond any assumption on the ultraviolet asymptotics, we show that if our physical requirements are enough to ensure the uniqueness of the quantization, up to unitary equivalence, then the time-dependent scaling in the definition of the annihilation and creation-like variables of our reference complex structure is the only one allowed. Finally, in Sec. \ref{sec:conclu} we summarize and further discuss our results, we explore possible generalizations of them, and we briefly comment on their applications, both in cosmology and in condensed matter systems.

\section{The classical model}
\label{sec:Model}

We consider the Einstein-Dirac system studied in Ref. \cite{H-D}, namely, a purely inhomogeneous massive Dirac field $\Psi$ propagating in a closed FRW cosmology with metric 
\begin{align}
ds^{2}=e^{2\alpha(\eta)}(-d\eta^{2}+d\Omega_{3}^{2}).
\end{align}
Here $\eta$ denotes conformal time, $d\Omega_{3}^{2}$ is the metric on the unit three-sphere $S^{3}$, and $\exp{(\alpha)}$ is the scale factor. Note that we have set the lapse as corresponding to conformal time. The Dirac spinor is taken in the Weyl representation of the constant Dirac matrices, $\gamma^{a}$, with $a=0,1,2,3$ (see Appendix \ref{appendixA1}). Thus, $\Psi$ can be described by an independent pair of  two-component spinors, $\phi^{A}$ and ${\bar{\chi}}_{A'}$, each of them with well defined and opposite chirality. We are using the index notation $A,B,...=0,1$ and $A',B',...=0',1'$ to denote the Grassmann variables \cite{Berezin} forming the two-component spinors, and we will follow the spinor conventions of Ref.  \cite{H-D}. In particular, we raise and lower spinor indices using the alternating spinors $\epsilon^{AB}$, $\epsilon_{AB}$, $\epsilon^{A'B'}$, $\epsilon_{A'B'}$, each of which is given by the matrix 
\begin{align}
\begin{pmatrix}
0 & 1 \\ -1 & 0
\end{pmatrix}.\nonumber
\end{align}
For example,  $\phi_{A}=\phi^{B}\epsilon_{BA}$ and ${\bar{\chi}}^{A'}=\epsilon^{A'B'}{\bar{\chi}}_{B'}$.

\subsection{Mode decomposition}

The spinors $\phi_{A}$ and ${\bar{\chi}}_{A'}$ can be expanded in the bases of spinor harmonics on $S^{3}$ provided, respectively, by the eigenmodes of the Dirac operators 
\begin{align}
-i n_{AA'}\,e^{BA'j}\,{}^{(3)}\!D_{j} \qquad {\rm and} \qquad -i n_{AA'}\,e^{AB'j}\,{}^{(3)}\!D_{j}.
\end{align} 
Here, $j=1,2,3$ is a spatial index, $e^{AA'j}$ is the spinor version of the triad on $S^{3}$, ${}^{(3)}\!D_{j}$ is the covariant derivative operator associated with the $SU(2)$ spin connection on $S^{3}$, and $n^{AA'}$ is the spinor version of the (unit, timelike, future-directed Lorentzian) normal to $S^{3}$.

These complete sets of spinor harmonics were already employed in Ref. \cite{H-D}. For the spinors with the chirality of $\phi_{A}$, a basis is given by the set of eigenmodes $\rho_{A}^{np}$ and ${\bar{\sigma}}^{np}_{A}$ that verify
\begin{align}\label{dirac1}
-in_{AA'}\,e^{BA'j}\,{}^{(3)}\!D_{j}\rho^{np}_{B}=\frac{\omega_{n}}{2}\rho^{np}_{A}, \qquad
-in_{AA'}\,e^{BA'j}\,{}^{(3)}\!D_{j}{\bar{\sigma}}^{np}_{B}=-\frac{\omega_{n}}{2}{\bar{\sigma}}^{np}_{A}.
\end{align}
Here $\omega_{n}=n+3/2$, with $n\in\mathbb{N}$. The degeneracy of each eigenspace is 
\begin{align}
g_n=(n+1)(n+2)=\omega_n^2-\frac{1}{4}.
\end{align} 
This degeneracy is accounted for by the label $p=1,...,g_n$. Analogously, a complete set of spinor harmonics on $S^{3}$ with the opposite chirality is provided by the Hermitian conjugate of the previous set, namely, by ${\bar{\rho}}^{np}_{A'}$ and $\sigma_{A'}^{np}$, modes that solve the equations 
\begin{align}\label{dirac2}
-in_{AA'}\,e^{AB'j}\,{}^{(3)}\!D_{j}{\bar{\rho}}^{np}_{B'}=-\frac{\omega_{n}}{2}{\bar{\rho}}^{np}_{A'},\qquad
-in_{AA'}\,e^{AB'j}\,{}^{(3)}\!D_{j}\sigma^{np}_{B'}=\frac{\omega_{n}}{2}\sigma^{np}_{A'}.
\end{align}
Indeed, these eigenmodes of the Dirac operators provide orthogonal bases of spinor harmonics on $S^{3}$, as they verify the orthogonality relations \cite{H-D}
\begin{align}
\label{inner-product1}
\int d\mu \, \rho_{A}^{np}n^{AA'}\sigma_{A'}^{mq}=0, \qquad &\int d\mu \, {\bar{\rho}}_{A'}^{np}n^{AA'}{\bar{\sigma}}_{A}^{mq}=0,\\
\label{inner-product}
\int d\mu \, \rho_{A}^{np}n^{AA'}{\bar{\rho}}_{A'}^{mq}=\delta^{nm}\delta^{pq}, \qquad
&\int d\mu \,{\bar{\sigma}}_{A}^{np}n^{AA'}\sigma_{A'}^{mq}=\delta^{nm}
\delta^{pq},
\end{align}
for all $n$, $p$, $m,$ and $q$,
where, using the same notation as Ref. \cite{H-D}, $d\mu$ denotes the volume element on $S^{3}$. 

In conclusion, the two-component spinors that describe the Dirac field, and their Hermitian conjugates, adopt the following expansion:
\begin{align}\label{harm1}
\phi_A(x)=&\frac{e^{-3\alpha(\eta)/2}}{2\pi}\sum_{npq}\breve{\alpha}_{n}^{pq}[m_{np}(\eta)\rho_{A}^{nq}(\textbf{x})+{\bar{r}}_{np}(\eta){\bar\sigma}_{A}^{nq}(\textbf{x})],
\\\label{harm2}
{\bar\phi}_{A'}(x)=&\frac{e^{-3\alpha(\eta)/2}}{2\pi}\sum_{npq}\breve{\alpha}_{n}^{pq}[{\bar{m}}_{np}(\eta){\bar\rho}_{A'}^{nq}(\textbf{x})+r_{np}(\eta)\sigma_{A'}^{nq}(\textbf{x})],
\\ \label{harm3}
\chi_A(x)=&\frac{e^{-3\alpha(\eta)/2}}{2\pi}\sum_{npq}\breve{\beta}_{n}^{pq}[s_{np}(\eta)\rho_{A}^{nq}(\textbf{x})+{\bar{t}}_{np}(\eta){\bar\sigma}_{A}^{nq}(\textbf{x})],
\\ \label{harm4}
{\bar\chi}_{A'}(x)=&\frac{e^{-3\alpha(\eta)/2}}{2\pi}\sum_{npq}\breve{\beta}_{n}^{pq}[{\bar{s}}_{np}(\eta){\bar\rho}_{A'}^{nq}(\textbf{x})+t_{np}(\eta)\sigma_{A'}^{nq}(\textbf{x})],
\end{align}
with
\begin{align}
\sum_{npq}:=\sum_{n=0}^{\infty}\sum_{p=1}^{g_n}\sum_{q=1}^{g_n},\nonumber
\end{align}
and where the anticommuting nature of the spinors is captured by the Grassmann variables $m_{np}$, $r_{np}$, $t_{np}$, $s_{np}$ (and their complex conjugates). Here, the constant coefficients $\breve{\alpha}_{n}^{pq}$ and $\breve{\beta}_{n}^{pq}$ are included for convenience, in order to avoid couplings between different values of $p$ when introducing these expansions in the Einstein-Dirac action. They can be regarded as the coefficients of two real matrices $\breve{\alpha}_{n}$ and $\breve{\beta}_{n}$, each of dimension $g_n$.  These matrices are block-diagonal, with blocks given by
\begin{align}
\begin{pmatrix}
1 & 1 \\ 1 & -1
\end{pmatrix}\qquad \text{and} \qquad \begin{pmatrix}
1 & -1 \\ -1 & -1
\end{pmatrix}\nonumber
\end{align}
for $\breve{\alpha}_{n}$ and $\breve{\beta}_{n}$, respectively. 

Let us notice that the fields have been scaled with a time-dependent factor $\exp{(3\alpha/2)}$ in these expansions. This scaling, or more generically, the one obtained by multiplying the fields by the fourth root of the determinant of the metric of the spatial sections, is often present in the Hamiltonian formulation of fermion fields in globally hyperbolic spacetimes \cite{T-N}. The reason underlying this fact can be understood as the way to obtain Dirac brackets for the fields that are free of any background structure, once the second class constraints of the system have been eliminated \cite{Dirac}. Actually, in the particular case under discussion, we will see in Sec. \ref{sec:uniqueness} that this time-dependent scaling is needed in order to allow for a unitary implementation of the dynamics of the field in the quantum theory.

\subsection{Fermion dynamics}

Employing the above decomposition, we may pass from a study of the spatial dependence to a spectral analysis in terms of modes that decouple dynamically. In fact, after introducing the harmonic expansions \eqref{harm1}-\eqref{harm4}, the dynamics of the fermion field deduced from the Einstein-Dirac action can be summarized in the following set of first-order equations, that are just the components of the (spatially integrated) Dirac equations of the model \cite{H-D}:
\begin{align}\label{1order}
x_{np}'=i\omega_{n}x_{np}-ime^{\alpha}\bar{y}_{np}, \qquad y_{np}'=i\omega_{n}y_{np}+ime^{\alpha}\bar{x}_{np},
\end{align}
together with their complex conjugates. Here $m$ is the mass of the fermion field, and the prime stands for the derivative with respect to conformal time. Besides, we are adopting the notation $(x_{np},y_{np})$ to denote indistinctly any of the sets of modes pairs  $(m_{np},s_{np})$ or $(t_{np},r_{np})$, as they both obey the same dynamics. These equations can be combined into a second-order form for all modes $\{x_{np},y_{np}\}:=  \{z_{np}\}$ which reads
\begin{align}\label{2order}
z_{np}''=\alpha 'z_{np}'-(\omega_{n}^2+m^{2}e^{2\alpha}+i\omega_{n}\alpha ')z_{np},
\end{align}
and into the corresponding complex conjugate equation for $\{{\bar x}_{np},{\bar y}_{np}\}:= \{{\bar z}_{np}\}$. If a complete Hamiltonian analysis of the Einstein-Dirac action is performed, one finds that the only non-vanishing Dirac brackets $\{\;,\;\}$ of the field variables are \cite{H-D}
\begin{align}\label{canonicalbrackets}
\{x_{np},\bar{x}_{np}\}=-i, \qquad \{y_{np},\bar{y}_{np}\}=-i.
\end{align}
Let us notice that these brackets are symmetric owing to the anticommutativity of the Grassmann variables considered here. Hence, upon quantization, they become anticommutators of the corresponding operators \cite{Casal}.

\subsection{Annihilation and creation-like variables}
\label{reference}

We will now introduce our preferred choice of annihilation and creation-like variables (both for particles and antiparticles). Here, and in what follows, we adopt the convention that the concepts of particle and antiparticle are associated, respectively, with the {\it positive} and (the complex conjugate of the) {\it negative} frequency contributions to the solutions of the Dirac equations (as it is usually done in the case of Minkowski spacetime). In Sec. \ref{J} we will make precise what we mean by positive and negative frequency contributions in the non-stationary case under study.

We call $a^{(x,y)}_{np}$ and $b^{(x,y)}_{np}$ the annihilation-like variables of the particles and antiparticles, respectively, associated with either the pair $(m_{np},s_{np})$ or $(t_{np},r_{np})$. We choose them as follows:
\begin{align}\label{refcs}
a^{(x,y)}_{np}=\frac{me^{\alpha}}{2\omega_{n}}x_{np}+\sqrt{1-\frac{m^{2}e^{2\alpha}}{4\omega_{n}^{2}}}\bar{y}_{np}, \qquad b^{(x,y)}_{np}=\sqrt{1-\frac{m^{2}e^{2\alpha}}{4\omega_{n}^{2}}}\bar{x}_{np}-\frac{me^{\alpha}}{2\omega_{n}}y_{np}.
\end{align}
The creation-like variables $a_{np}^{(x,y)\dagger}:=\bar{a}_{np}^{(x,y)}$ and $b_{np}^{(x,y)\dagger}:=\bar{b}_{np}^{(x,y)}$ are their complex conjugates. It is straightforward to check that these variables indeed satisfy the Dirac brackets characteristic of annihilation and creation-like variables for particles and antiparticles, namely
\begin{align}\label{CACR}
\{a^{(x,y)}_{np},a^{(x,y)\dagger}_{np}\}=
\{b^{(x,y)}_{np},b^{(x,y)\dagger}_{np}\}=-i,\qquad \{a^{(x,y)}_{np},b^{(x,y)}_{np}\}=0.
\end{align}
This fact guarantees the invertibility of the relation between such variables and the set of modes $\{x_{np},y_{np},\bar{x}_{np},\bar{y}_{np}\}$. Let us notice that, upon substitution of this inverse relation in the harmonic expansions \eqref{harm1} and \eqref{harm4}, we get that the Dirac spinor $\Psi$ is described by the two-component spinors
\begin{align}
\phi_A=&\frac{e^{-3\alpha/2}}{2\pi}\sum_{npq}\breve{\alpha}_{n}^{pq}\Bigg[\left(\frac{me^{\alpha}}{2\omega_{n}}a_{np}^{(m,s)}+\sqrt{1-\frac{m^{2}e^{2\alpha}}{4\omega_{n}^{2}}}b_{np}^{(m,s)\dagger}\right)\rho_{A}^{nq}\nonumber \\ &+\left(\sqrt{1-\frac{m^{2}e^{2\alpha}}{4\omega_{n}^{2}}}a_{np}^{(t,r)}-\frac{me^{\alpha}}{2\omega_{n}}b_{np}^{(t,r)\dagger}\right){\bar\sigma}_{A}^{nq}\Bigg],
\\
{\bar\chi}_{A'}=&\frac{e^{-3\alpha/2}}{2\pi}\sum_{npq}\breve{\beta}_{n}^{pq}\Bigg[\left(\sqrt{1-\frac{m^{2}e^{2\alpha}}{4\omega_{n}^{2}}}a_{np}^{(m,s)}-\frac{me^{\alpha}}{2\omega_{n}}b_{np}^{(m,s)\dagger}\right){\bar\rho}_{A'}^{nq}\nonumber \\ &+\left(\frac{me^{\alpha}}{2\omega_{n}}a_{np}^{(t,r)}+\sqrt{1-\frac{m^{2}e^{2\alpha}}{4\omega_{n}^{2}}}b_{np}^{(t,r)\dagger}\right)\sigma_{A'}^{nq}\Bigg].
\end{align}
Therefore, for our choice of variables, the fermion field presents specific and different time-dependent scalings in its particle and antiparticle parts, scalings which are different as well for each of the two chiralities. This feature will be relevant when analyzing the quantization.

As we will explain in the next section, our choice of annihilation and creation-like variables is equivalent to choosing a particular complex structure, and hence, a particular Fock quantization. We will call reference complex structure and reference quantization the ones determined by the set of variables that we have introduced above.

\section{Invariance of the vacuum and unitary evolution}
\label{sec:Unitarity}

From a physical point of view, we want to restrict our attention to those Fock representations that satisfy the criteria put forward in Refs. \cite{gowdy,spheres,compact,torus}, namely: i) invariance of the vacuum under the symmetries of the evolution equations, and ii) unitary implementability of the dynamics in the quantum theory. Let us first review the notion of complex structure and how it characterizes the quantum representation (for more details see e.g. \cite{bratteli}).

\subsection{Complex structure and one-particle Hilbert space}

\label{J}

Let ${S}=\{\Psi\}$ be the complex vector space of Dirac spinors, solutions of the Dirac equation on the cosmological background under study. Let us notice that this vector space is isomorphic to the space of initial data for the Dirac equation, given the well posedness of the Cauchy problem for such equation in globally hyperbolic spacetimes \cite{Dimock}. The linear space ${S}$ is naturally equipped with the inner product \cite{Dimock}:
\begin{align}\label{inner}
(\Psi_1,\Psi_2)_S=\int d\tilde\mu\, \Psi^{+}_1 n^{\nu}e_{\nu}^{a}\gamma_{a} \Psi_2,
\end{align}
where the right-hand side is evaluated at a certain and arbitrary time, and $d\tilde\mu$ is the integration measure on the spatial sections [i.e., $d\tilde\mu=\exp{(3\alpha)} d\mu$ in our model]. We have used here the notation $\Psi^{+}=\Psi^{\dagger}\gamma_{0}$ to denote the adjoint Dirac spinor. Besides, $n^{\nu}$ are the spacetime components of the (unit, timelike, future-directed Lorentzian) normal to the spatial sections (with $\nu=0,1,2,3$), and $e^{a}_{\nu}$ is the tetrad. This inner product can be seen to be conserved under the evolution of the solutions of the Dirac equation \cite{Dimock}. Furthermore, one can indeed prove that it is positive definite. In particular, given the Weyl representation used here, it can be checked that
\begin{align}
(\Psi_1,\Psi_2)_S=\int d\tilde\mu\, \chi_{1A}n^{AA'}\bar{\chi}_{2A'}+\int d\tilde\mu\, \bar{\phi}_{1A'}n^{AA'}\phi_{2A},
\end{align}
which is clearly positive definite given the orthonormality relations \eqref{inner-product1} and \eqref{inner-product}. 

On the other hand, let $\bar{S}$ be the complex conjugate of $S$, with inner product given by the complex conjugate of \eqref{inner}. A complex structure $J: S\rightarrow S$ is then a real linear map with the property $J^2=-I$, and such that it leaves the inner product invariant, that is $(J\Psi_1,J\Psi_2)_S=(\Psi_1,\Psi_2)_S$. Any complex structure $J$ defines a splitting $S=S_J^+\oplus S_J^-$ of $S$ into two mutually complementary subspaces $S^\pm_J=(S\mp iJS)/2$. Indeed, it is easy to check that $S_J^+$ and $S_J^-$ are orthogonal with respect to the considered inner product. Note that $S_J^\pm$ are the eigenspaces of $J$, with eigenvalue $\pm i$. They provide the decomposition of $\Psi$ into what we call the positive and negative frequency contributions, mentioned at the beginning of Sec. \ref{reference}. So, our convention is to assign $S_J^+$ to the space of particles, and $S_J^-$ to the space of antiparticles.

Analogously, we can define the complex structure $J$ as a linear map on $\bar{S}$. It then induces the particle-antiparticle splitting $\bar{S}=(\bar{S}_J)^+\oplus(\bar{S}_J)^-$, with 
\begin{align}
(\bar{S}_J)^\pm=\frac{1}{2}(\bar{S}\mp iJ\bar{S})=\overline{(S_J^\mp)}.
\end{align} 
Now, we define  {\it the one-particle Hilbert space of particles} as the completion of $S_J^+$ in the inner product \eqref{inner}, and {\it the one-particle Hilbert space of antiparticles} as the completion of $\overline{(S^-_J)}$. We denote these Hilbert spaces by $H_J^+$ and $\overline{H_J^-}$, respectively. Then, the one-particle Hilbert space of the quantum theory associated with the complex structure $J$, from which one constructs the antisymmetric Fock space, is taken to be the Hilbert space $\mathcal H_J=H_J^+\oplus \overline{H_J^-}$.

From this explanation, it is now clear that different complex structures define different concepts of particle and antiparticle, with their respective annihilitation and creation-like variables, and hence different one-particle Hilbert spaces. Here is where the ambiguity in the Fock quantization of the Dirac field resides. In practice, the choice of a complex structure is the same as the choice of a set of classical annihilitation and creation-like variables, to be quantized as  annihilitation and creation operators.

\subsection{Invariant complex structures}

We will now determine the general form that a complex structure must have to be invariant under the symmetries of the Dirac equations of motion \eqref{1order}. Of course, these symmetries include the isometry group of the considered spatial sections, namely $SO(4)$, that in particular preserves the measure on $S^3$.

We have seen that all pairs of modes $(m_{np}, \bar{s}_{np})$ and $(t_{np},\bar{r}_{np})$ become decoupled in the evolution dictated by the Dirac equations, for all values of the labels $n$ and $p$. As a consequence, any complex structure $J$ that shares the symmetries of the dynamics must be block-diagonal in the basis provided by the union of these modes, the only non-trivial blocks being of dimension 2 and mixing the commented pairs. Furthermore, the dynamical equations depend only on the (norm of the) eigenvalue of the Dirac operator, and therefore they are invariant under the interchange of any of the mentioned pairs if the corresponding Dirac eigenvalue is the same. In other words, the group of symmetries of the field equations includes all transformations performing any possible interchange of dynamically decoupled pairs in the different eigenspaces of the Dirac operator that are characterized by the integer $n$. It is then straightforward to realize (for instance, using Schur's lemma \cite{Reps}) that the diagonal $2\times 2$ blocks that define the invariant complex structures can only depend on the number $n$, but not on the label $p$ that lists the various modes with the same Dirac eigenvalue, nor on the consideration of pairs of the type $(m_{np}, \bar{s}_{np})$ or $(t_{np},\bar{r}_{np})$. Finally, since the evolution equations are different for different values of $n$, it is clear that there are no more restrictions on the complex structures owing to symmetries of the dynamics.

Summarizing, an invariant complex structure, that commutes with the action of the group of symmetries of the evolution equations, diagonalizes into $2\times 2$ blocks. These blocks can at most mix the modes $(m_{np}, \bar{s}_{np})$ or $(t_{np},\bar{r}_{np})$ with the same value of $p$, and are all equal for modes associated with the same eigenvalue of the Dirac operator (in norm). In this sense, invariant complex structures are totally determined by a series of $2 \times 2$ matrices labelled by the integer $n\in \mathbb{N}$.

Notice that, for each label $p$ of the degeneracy associated with a given $n$, there correspond two particle annihilation-like variables and two antiparticle ones, given the two pairs of dynamically decoupled modes. These two degrees of freedom, additional to the charge parity of the particle, just account for the two possible helicities of a Dirac fermion.

\subsection{Conditions for unitary dynamics}

In the light of the previous discussion, we conclude that the particle and antiparticle annihilation and creation-like variables corresponding to our class of invariant complex structures are such that, at any time $\eta$,
\begin{align}\label{blockcs1}
\begin{pmatrix} a_{np}^{(x,y)} \\ b^{(x,y)\dagger}_{np} \\ a^{(x,y)\dagger}_{np} \\ b_{np}^{(x,y)}\end{pmatrix}_{\!\!\eta}=\begin{pmatrix}f_{1}^ {n}(\eta) & f_{2}^{n}(\eta) & 0 & 0 \\ g_{1}^ {n}(\eta) & g_{2}^{n}(\eta) & 0 & 0 \\ 0 & 0 & \bar{f}^{n}_{1}(\eta) & \bar{f}^{n}_{2}(\eta) \\ 0 & 0 & \bar{g}^{n}_{1}(\eta) & \bar{g}^{n}_{2}(\eta)\end{pmatrix}\begin{pmatrix} x_{np} \\ \bar{y}_{np} \\ \bar{x}_{np} \\ y_{np}\end{pmatrix}_{\!\!\eta},
\end{align}
where $(x_{np},y_{np})$ stands again either for $(m_{np},s_{np})$ or $(t_{np},r_{np})$, and the subindex $\eta$ in column-vectors stands for evaluation at that value of the conformal time. The time-dependent functions $f_{l}^{n}$ and $g_{l}^{n}$ (here, and in what follows, $l=1,2$), together with  their complex conjugates $\bar{f}^{n}_{l}$ and $\bar{g}^{n}_{l}$, satisfy the relations
\begin{align}\label{sympl}
|f_{1}^{n}|^{2}+|f_{2}^{n}|^{2}=1, \qquad |g_{1}^{n}|^{2}+|g_{2}^{n}|^{2}=1, \qquad f_{1}^{n}\bar{g}^{n}_{1}+f_{2}^{n}\bar{g}^{n}_{2}=0,
\end{align}
as required by demanding Eq. \eqref{CACR}. These relations guarantee the invertibility of Eq. \eqref{blockcs1}, and, conveniently combined, they allow us to write
\begin{align}\label{fgrel}
g_{1}^{n}=\bar{f}_{2}^{n}e^{iG^{n}}, \qquad g_{2}^{n}=-\bar{f}^{n}_{1}e^{iG^{n}},
\end{align}
with $G^{n}$ a certain phase. In particular, we have
\begin{align}\label{fgdif}
f_{1}^{n}g^{n}_{2}-g^{n}_{1}f^{n}_{2}=-e^{iG^{n}}.
\end{align}
It then follows that only one of the four functions $\{f_{l}^{n},g_{l}^{n}\}$, together with two additional phases for each $n$, suffice to fully characterize the considered complex structure.

We proceed to analyze the Bogoliubov transformations that the fermion dynamics of the system induces on this class of invariant complex structures. In order to do so, let us recall that both modes $x_{np}$ and $y_{np}$ obey the same second-order differential equation, given by Eq. \eqref{2order}, which furthermore it is the same for all the modes with the same value of $n$, regardless of the label $p$. Hence, all of them are linear combinations of two complex and independent solutions that we will call $\exp[{i\Theta^{1}_{n}(\eta)}]$ and $\exp[{-i\Theta^{2}_{n}(\eta)}]$. In general, neither $\Theta^{1}_{n}$ nor $\Theta^{2}_{n}$ will be real owing to the fact that the differential equation is complex. Let us set the following generic initial conditions at a specific time $\eta_{0}$:
\begin{align}\label{ic}
\Theta^{l}_{n}(\eta_{0})=\Theta_{n,0}^{l},  \qquad (\Theta^{l}_{n})'(\eta_{0})=\Theta^{l}_{n,1}.
\end{align}
We also call $\Omega^{1}_{n,0}=\exp({i\Theta^{1}_{n,0}})$ and $\Omega^{2}_{n,0}=\exp({-i\Theta^{2}_{n,0}})$ to describe the initial conditions for the independent solutions at the instant $\eta_{0}$. Notice that the initial conditions on $\Theta^{l}_{n}$ and their derivatives are related to the initial conditions $x^{0}_{np}$ and $y^{0}_{np}$ on the modes, and on their complex conjugates, via the Dirac equations \eqref{1order}. Taking this into account, one can easily derive the expression for the fermion modes at any time $\eta$, in terms of the two independent solutions of Eq. \eqref{2order} and their initial conditions:
\begin{align}\label{evmodes}
&x_{np}(\eta)=\left[\Delta_{n}^{2}e^{i\Theta^{1}_{n}(\eta)}+\Delta^{1}_{n}e^{-i\Theta_{n}^{2}(\eta)}\right]x^{0}_{np}-\left[\Gamma^{1}_{n}e^{i\Theta_{n}^{1}(\eta)}-\Gamma_{n}^{2}e^{-i\Theta_{n}^{2}(\eta)}\right]\bar{y}^{0}_{np}\nonumber\\ &
y_{np}(\eta)=\left[\Delta_{n}^{2}e^{i\Theta^{1}_{n}(\eta)}+\Delta^{1}_{n}e^{-i\Theta_{n}^{2}(\eta)}\right]y^{0}_{np}+\left[\Gamma^{1}_{n}e^{i\Theta_{n}^{1}(\eta)}-\Gamma_{n}^{2}e^{-i\Theta_{n}^{2}(\eta)}\right]\bar{x}^{0}_{np},
\end{align}
and we have introduced the constants
\begin{align}\label{icconsts}
\Delta^{1}_{n}=\frac{\Theta_{n,1}^{1}-\omega_{n}}{\Omega^{2}_{n,0}(\Theta^{1}_{n,1}+\Theta^{2}_{n,1})}, \qquad \Delta^{2}_{n}=\frac{\Theta_{n,1}^{2}+\omega_{n}}{\Omega^{1}_{n,0}(\Theta^{1}_{n,1}+\Theta^{2}_{n,1})}, \end{align}
and \begin{align}\Gamma_{n}^{l}=\frac{m e^{\alpha_0}}{\Omega^{l}_{n,0}(\Theta^{1}_{n,1}+\Theta^{2}_{n,1})},\label{icconsts2}
\end{align}
where $\alpha_{0}=\alpha(\eta_{0})$.

With these expressions at hand, and inverting Eq. \eqref{blockcs1}, we can obtain the form of the Bogoliubov transformation $B_{n}(\eta,\eta_{0})$ that implements the dynamics from the initial time $\eta_{0}$ to any other time $\eta$. We obtain
\begin{align}\label{bog}
\begin{pmatrix} a_{np}^{(x,y)} \\ b^{(x,y)\dagger}_{np} \\ a^{(x,y)\dagger}_{np} \\ b_{np}^{(x,y)}\end{pmatrix}_{\!\!\eta}=B_{n}(\eta,\eta_{0})\begin{pmatrix} a_{np}^{(x,y)} \\ b^{(x,y)\dagger}_{np} \\ a^{(x,y)\dagger}_{np} \\ b_{np}^{(x,y)}\end{pmatrix}_{\!\!\eta_{0}},
\end{align}
with
\begin{align}
B_{n}=\begin{pmatrix} \mathcal{B}_{n} & 0 \\ 0 & \mathcal{\bar{B}}_{n} \end{pmatrix}, \qquad \mathcal{B}_{n}=\begin{pmatrix} \alpha_{n}^{f} & \beta_{n}^{f} \\ \beta_{n}^{g} & \alpha_{n}^{g} \end{pmatrix},
\end{align}
and with alpha and beta coefficients given by
\begin{align}
\alpha_{n}^{h}=&\frac{1}{h_{1}^{n,0}k_{2}^{n,0}-h_{2}^{n,0}k_{1}^{n,0}}
\bigg\{\big[\Delta_{n}^{2}k_{2}^{n,0}+\Gamma^{1}_{n}k_{1}^{n,0}\big]h_{1}^{n}e^{i\Theta_{n}^{1}}+\big[\Delta_{n}^{1}k_{2}^{n,0}-\Gamma^{2}_{n}k_{1}^{n,0}\big]h_{1}^{n}e^{-i\Theta_{n}^{2}}
\nonumber \\ 
-&\big[\bar{\Delta}_{n}^{1}k_{1}^{n,0}+\bar{\Gamma}^{2}_{n}k_{2}^{n,0}\big]h_{2}^{n}e^{i\overline{\Theta}_{n}^{2}}-\big[\bar{\Delta}_{n}^{2}k_{1}^{n,0}-\bar{\Gamma}^{1}_{n}k_{2}^{n,0}\big]h_{2}^{n}e^{-i\overline{\Theta}_{n}^{1}}\bigg\}, \\
\beta_{n}^{h}=&-\frac{1}{h_{1}^{n,0}k_{2}^{n,0}-h_{2}^{n,0}k_{1}^{n,0}}
\bigg\{\big[\Delta_{n}^{2}h_{2}^{n,0}+\Gamma^{1}_{n}h_{1}^{n,0}\big]h_{1}^{n}e^{i\Theta_{n}^{1}}+\big[\Delta_{n}^{1}h_{2}^{n,0}-\Gamma^{2}_{n}h_{1}^{n,0}\big]h_{1}^{n}e^{-i\Theta_{n}^{2}} \nonumber \\ 
-&\big[\bar{\Delta}_{n}^{1}h_{1}^{n,0}+\bar{\Gamma}^{2}_{n}h_{2}^{n,0}\big]h_{2}^{n}e^{i\overline{\Theta}_{n}^{2}}-\big[\bar{\Delta}_{n}^{2}h_{1}^{n,0}-\bar{\Gamma}^{1}_{n}h_{2}^{n,0}\big]h_{2}^{n}e^{-i\overline{\Theta}_{n}^{1}}\bigg\}.\label{beta}
\end{align}
Again, overbarred symbols denote complex conjugates. Here, $\{h,k\}:=\{f,g\}$ as a set, with $h$ being equal to either $f$ or $g$ and $k$ being the complementary of $h$. We have omitted the dependence of these functions on $\eta$ to alleviate the notation, and distinguished evaluation at $\eta_{0}$ with the superscript $0$ (preceded by a comma). 

Let us consider a complex structure $J$ on the space of solutions, and let us call $J_{\eta_0}$ the complex structure corresponding to it on the space of initial conditions at a given time $\eta_0$ (via the isomorphism between the two spaces). Let us also call $J_\eta$ the complex structure on this very space of initial conditions obtained from $J_{\eta_0}$ by the transformation provided by the dynamical evolution from $\eta_0$ to the time $\eta$. Both complex structures are related precisely by the Bogoliubov transformation studied above, defined by the blocks $B_{n}(\eta,\eta_{0})$. In this framework, it is clear that the dynamics admits a unitary implementation in the Fock representation determined by $J_{\eta_0}$ if and only if the representations determined by $J_\eta$ and $J_{\eta_0}$ are unitarily equivalent for all the allowed values of $\eta$. This is equivalent to demand that the operator $J_\eta-J_{\eta_0}$ be Hilbert-Schmidt (on the one-particle Hilbert space $\mathcal H_{J_{\eta_0}}$) \cite{Shale}. In turn, this condition is true if and only if the beta coefficients of the Bogoliubov transformation relating $J_\eta$ with $J_{\eta_0}$ are square summable for all values of $\eta$, namely if and only if the sum
\begin{align}
\sum_{n}g_n(|\beta^{f}_{n}|^{2}+|\beta^{g}_{n}|^{2})
\end{align}
is convergent at all times. Notice that we have taken into account the degeneracy $g_n$ associated with each block labeled by $n$. Now, since all the terms of the sum are positive, the convergence amounts to demand that
\begin{align}\label{ucond}
\sum_{n}g_{n}|\beta_{n}^{f}|^{2}<\infty, \qquad \sum_{n}g_{n}|\beta_{n}^{g}|^{2}<\infty.
\end{align}
These conditions impose restrictions in the ultraviolet regime on the beta coefficients, regime given by the sector of asymptotically large values of $n$, or equivalently of large absolute values of the Dirac eigenvalues $\pm\omega_n/2$. Indeed, if we assume e.g. that the norm of the beta coefficients admits a Laurent series in the eigenvalue $\omega_{n}$, at least up to terms of order $\omega_n^{-2}$, conditions \eqref{ucond} are attained if and only if $|\beta^{f}_{n}|$ and  $|\beta^{g}_{n}|$ are negligible in comparison with $\omega_n^{-1}$ in the considered asymptotic regime, since we recall that $g_n=\omega_{n}^{2}-1/4$.

In order to select a class of complex structures that satisfy conditions \eqref{ucond}, and in particular to check whether the reference quantization introduced in Sec. \ref{sec:Model} satisfies them (as well as if the same happens with the quantization contemplated in Ref. \cite{H-D}), we will now analyze the dynamical behavior of the beta coefficients \eqref{beta}. A detailed asymptotic study of the solutions of the dynamical equation \eqref{2order} can be found in Appendix \ref{appendixA}. From such analysis it follows that, given some suitable initial conditions and some mild requirements on the time dependence of the background scale factor, the dynamical beta coefficients can be written as
\begin{align}\label{beta2}
\beta_{n}^{h}=&\frac{1}{h_{1}^{n,0}k_{2}^{n,0}-h_{2}^{n,0}k_{1}^{n,0}}
\Bigg\{\left[-h_{1}^{n}\bigg(h_{2}^{n,0}+\frac{\Gamma^{(\omega)}_{n}}{\omega_{n}}h_{1}^{n,0}\bigg)e^{i\int \Lambda^{1}_{n}}+\frac{\overline{\Gamma}^{(\omega)}_{n}}{\omega_{n}}h_{2}h_{2}^{n,0}e^{\Delta\alpha}e^{i\int\overline{\Lambda}^{2}_{n}}\right]
e^{i\omega_{n}\Delta\eta}\nonumber\\&
+\left[h_{2}^{n}\bigg(h_{1}^{n,0}-\frac{\overline{\Gamma}^{(\omega)}_{n}}{\omega_{n}}h_{2}^{n,0}\bigg)e^{-i\int \overline{\Lambda}^{1}_{n}}+\frac{\Gamma^{(\omega)}_{n}}{\omega_{n}}h_{1}h_{1}^{n,0}e^{\Delta\alpha}e^{-i\int\Lambda^{2}_{n}}\right]
e^{-i\omega_{n}\Delta\eta}\Bigg\},
\end{align}
where $\Delta\alpha=\alpha-\alpha_{0}$, $\Delta\eta=\eta-\eta_{0}$, and we have defined the constant $\Gamma^{(\omega)}_{n}:=\omega_{n}\Gamma^{l}_{n}$, because $\Gamma^{1}_{n}=\Gamma^{2}_{n}$ with the given initial conditions. The integrals in this expression are in conformal time, in the interval $[\eta_{0},\eta]$, and $\Lambda^{j}_{n}$ are the time-dependent functions defined in Eq. \eqref{applaurent}, which have the property of being $\mathcal{O}(\omega_{n}^{-1})$ in the ultraviolet regime. 

Remarkably, the norm of these beta coefficients is invariant under the change of complex structure induced by the replacement
\begin{align}\label{symmetry}
h_{1}^{n}\longrightarrow \tilde{h}_{1}^{n}=-\bar{h}_{2}^{n} e^{i\delta_{n}}, \qquad h_{2}^{n}\longrightarrow \tilde{h}_{2}^{n}=\bar{h}_{1}^{n} e^{i\delta_{n}},
\end{align}
with $\delta_{n}$ any phase.
To check this symmetry, one first notices that the above replacement induces also the following one:
\begin{align}
k_{1}^{n}\longrightarrow \tilde{k}_{1}^{n}=-\bar{k}_{2}^{n}e^{-i\delta_{n}}e^{2iG^{n}}, \qquad k_{2}^{n}\longrightarrow \tilde{k}_{2}^{n}=\bar{k}_{1}^{n}e^{-i\delta_{n}}e^{2iG^{n}}.
\end{align}
Now, using Eq. \eqref{beta2}, we can check that under this change
\begin{align}
{\beta}^{h}_{n}\longrightarrow \tilde{\beta}^{h}_{n}=\bar{\beta}^{h}_{n}e^{i(\delta_n+\delta_{n}^{0}-2G^{n,0})},
\end{align}
and therefore $|\beta^{h}_{n}|= |\tilde{\beta}^{h}_{n}|,$ as we wanted to prove. Hence, we can conclude that, if conditions \eqref{ucond} hold for ${\beta}^{h}_{n}$, they hold as well for $\tilde{\beta}^{h}_{n}$. In other words, the complex structures related by the interchange $(h_{1}^{n},h_{2}^{n})\longleftrightarrow(\tilde h_{1}^{n},\tilde h_{2}^{n})$ both admit a unitary implementation of the dynamics. 

\subsection{Existence of unitary dynamics}

In the subsequent analysis, we will restrict all considerations to the class of invariant complex structures given by Eq. \eqref{blockcs1} and such that they display a particular ultraviolet behavior. Specifically, we will consider complex structures for which either $h_{1}^{n}= \mathcal{O}(\omega_{n}^{-1})$ or $h_{2}^{n}= \mathcal{O}(\omega_{n}^{-1})$ in the ultraviolet regime, and the corresponding next-to-leading order term is $\mathcal{O}(\omega_{n}^{-2})$. It suffices to focus on one of the two possibilities owing to the symmetry property $|\beta^{h}_{n}|= |\tilde{\beta}^{h}_{n}|$ under the interchange $(h_{1}^{n},h_{2}^{n})\longleftrightarrow(\tilde{h}_{1}^{n},\tilde{h}_{2}^{n})$, which in practice flips the role of $h_1^n$ and $h_2^n$. Thus, for concreteness, let us investigate the case with the asymptotic behavior $h_{1}^{n}=\mathcal{O}(\omega_{n}^{-1})$. We can write
\begin{align}\label{ultrav}
h_{1}^{n}=\frac{q^{n}}{\omega_{n}}+\mathcal{O}(\omega_{n}^{-2}), \qquad h_{2}^{n}=e^{iH^{n}}+\mathcal{O}(\omega_{n}^{-2}),\qquad q^{n}:=q e^{iQ^n},
\end{align}
where $Q^n$ is a phase and $q$ is a non-negative and $n$-independent  function of time. They contain the time dependence of $h^n_1$ at leading order. Besides, $H^{n}$ is another phase, and we have used the fact that $|h_2^n|= 1+\mathcal{O}(\omega_n^{-2})$, as the norms of $h_{1}^{n}$ and $h_{2}^{n}$ are related by Eq. \eqref{sympl}. Then, using Eq. \eqref{fgrel}, we can complete the characterization of the asymptotic behavior of the whole complex structure:
\begin{align}\label{ultravk}
k_{1}^{n}=e^{i\tilde{G}^{n}}e^{-iH^{n}}+\mathcal{O}(\omega_{n}^{-2}), \qquad k_{2}^{n}=-\frac{e^{i\tilde{G}^{n}}\bar{q}^{n}}{\omega_{n}}+\mathcal{O}(\omega_{n}^{-2}),
\end{align}
where $\tilde{G}^{n}$ is $G^{n}$ if $k=g$, whereas it equals $G^{n}+\pi$ if $k=f$. For this subclass of complex structures, half of the beta coefficients of the dynamical evolution ($\beta^{f}_{n}$ if $h=f$ or $\beta^{g}_{n}$ if $h=g$) adopt the following form:
\begin{align}
\beta^{h}_{n}=&\frac{e^{-i\tilde{G}^{n,0}}}{\omega_{n}}\bigg[\bigg(q^{n}-\frac{m}{2}e^{\alpha +iH^{n}}\bigg)e^{i\omega_{n}\Delta\eta+iH^{n,0}}
-\bigg(q^{n,0}-\frac{m}{2}e^{\alpha_{0} +iH^{n,0}}\bigg)e^{-i\omega_{n}\Delta\eta+iH^{n}}\bigg]+\mathcal{O}(\omega_{n}^{-2}).
\end{align}
Hence, the square summability condition for $\beta^{h}_{n}$ is satisfied if and only if
\begin{align}\label{unitcond1}
\sum_{n}\bigg|q^{n}(\eta)-\frac{m}{2}e^{\alpha(\eta)+iH^{n}(\eta)}\bigg|^{2}<\infty, 
\end{align}
for all times $\eta$, since $\exp({i\omega_{n}\Delta\eta})$ and $\exp({-i\omega_{n}\Delta\eta})$ are independent.\footnote{Owing to this and the fact that $\alpha$ is real, one concludes that the contribution of $\exp({\alpha+i\omega_{n}\Delta\eta+iH^n})$ (which would not be square summable by its own) may be compensated only by the term proportional to $q^n$. As a consequence, one finally arrives at the requirement \eqref{unitcond1}.} Then, in order to attain the convergence of the sum, we need to fix
\begin{align}\label{unitcond2}
q^{n}=\frac{me^{\alpha}}{2}e^{iH^{n}},
\end{align}
for all $n$ bigger than a certain $n_{0}\geq 0$.\footnote{Any subdominant term is absorbed in the additional contributions to the asymptotic expression \eqref{ultrav}.} 
Upon substitution of the condition \eqref{unitcond2} on the relations \eqref{ultrav}-\eqref{ultravk} for the asymptotic behavior, and taking into account the mentioned symmetry under the transformation \eqref{symmetry}, we get that the square summability of $\beta^{h}_{n}$ is satisfied if an only if
\begin{align}\label{hcoefs}
h_{l}^{n}=\frac{me^{\alpha}}{2\omega_{n}}e^{iH^{n}}+\mathcal{O}(\omega_{n}^{-2}), \qquad h_{\tilde{l}}^{n}=(-1)^{\tilde{l}}e^{iH^{n}}+\mathcal{O}(\omega_{n}^{-2}),
\end{align}
and
\begin{align}\label{kcoefs}
k_{l}^{n}=e^{-iH^{n}}e^{i\tilde{G}^{n}}+\mathcal{O}(\omega_{n}^{-2}), \qquad k_{\tilde{l}}^{n}=-(-1)^{\tilde{l}}\,\frac{me^{\alpha}}{2\omega_{n}}e^{-iH^{n}}e^{i\tilde{G}^{n}}+\mathcal{O}(\omega_{n}^{-2}).
\end{align}
Here $\{l,\tilde{l}\}=\{1,2\}$ as a set. We see that the pair $(k^{n}_{l},k_{\tilde{l}}^{n})$ has the same form as $(\tilde{h}^{n}_{l},\tilde{h}_{\tilde{l}}^{n})$ given by the transformation \eqref{symmetry}, taking $\delta_{n}=\tilde{G}^{n}-\pi$. Therefore, $|\beta^{k}_{n}|= |\tilde{\beta}^{h}_{n}|$, and the remaining beta coefficients, $\beta^{k}_{n}$, are automatically square summable when so are the coefficients $\beta^{h}_{n}$. In conclusion, there are no additional conditions, aside from Eq. \eqref{unitcond2}, to be fulfilled by the considered class of complex structures in order to admit a unitary implementation of the dynamics. 

As a summary, we have proven that the set of invariant complex structures characterized by annihilation and creation-like variables with an ultraviolet asymptotic behavior of the type \eqref{ultrav} allows for a unitary implementation of the dynamics if and only if the corresponding coefficient $q^{n}$ of the term of asymptotic order $\omega_{n}^{-1}$ is of the form \eqref{unitcond2} for all $n> n_{0}\geq 0$. 

One can easily check that the complex structure selected in Ref. \cite{H-D} is determined by annihilation and creation-like variables with the following real coefficients:
\begin{align}
f_{1}^{n}=\frac{m}{\sqrt{2\omega(\nu+\omega)}},\qquad f_{2}^{n}=\sqrt{\frac{\nu+\omega}{2\omega}}, \qquad g_{1}^{n}=f_{2}^{n}, \qquad g_{2}^{n}=-f_{1}^{n},
\end{align}
where (using the notation of that work), we have called $\nu=\omega_{n}\exp{(-\alpha)}$, and $\omega=\sqrt{\nu^{2}+m^{2}}$. In the ultraviolet sector, one then finds that
\begin{align}\label{theirs}
f_{1}^{n}=\frac{me^{\alpha}}{2\omega_{n}}+\mathcal{O}(\omega_{n}^{-3}),\qquad f_{2}^{n}=1+\mathcal{O}(\omega_{n}^{-2}).
\end{align}
This has motivated our choice of reference complex structure in Sec. \ref{sec:Model}, structure which is simply the truncation at leading order in the asymptotic expansion in $\omega_n$ of that of Ref. \cite{H-D}, namely, it is given by
\begin{align}\label{ours}
f_{1}^{n}=\frac{me^{\alpha}}{2\omega_{n}},\qquad f_{2}^{n}=\sqrt{1-|f_{1}^{n}|^2},\qquad g_{1}^{n}=f_{2}^{n}, \qquad g_{2}^{n}=-f_{1}^{n}.
\end{align}
Obviously, the choice \eqref{theirs} agrees with the result \eqref{hcoefs} (when $h=f$). Thus, both our reference Fock quantization and the one chosen in Ref. \cite{H-D} admit unitary implementable dynamics. Actually, this is the fundamental reason underlying the finite production of particles and antiparticles on the evolved vacuum found in Ref. \cite{H-D}.

In the following, we will call $J_\text{R}$ our reference complex structure [given by Eq. \eqref{ours}].

\section{Uniqueness of the representation and its scaling} 
\label{sec:uniqueness}

In the previous section we have derived the necessary and sufficient conditions that a specific set of complex structures need to satisfy for the dynamics to admit a unitary implementation in the quantum theory. However, the question of what is the relation between the associated Fock representations remains unanswered. This is actually an important issue, for if they were found to be unitarily inequivalent, the physical predictions resulting from these quantum theories would differ from one to another. We thus proceed now to analyze whether there exists unitary equivalence among the different complex structures constructed above, namely, those invariant complex structures with ultraviolet behavior of the kind \eqref{hcoefs}-\eqref{kcoefs} (so that they allow the unitary implementation of the dynamics).

\subsection{Unitary equivalence}

Let us consider two of such complex structures $J$ and $\tilde{J}$ on the space of initial data (at the given fixed time), provided with the algebra of Dirac brackets of anticommuting variables \cite{Dirac,Casal}. These complex structures will be characterized by two different sets of annihilation and creation-like variables, 
\begin{align}
\{a_{np}^{(x,y)},b_{np}^{(x,y)},a_{np}^{(x,y)\dagger},b_{np}^{(x,y)\dagger}\}\qquad {\rm and} \qquad \{\tilde{a}_{np}^{(x,y)},\tilde{b}_{np}^{(x,y)},\tilde{a}_{np}^{(x,y)\dagger},\tilde{b}_{np}^{(x,y)\dagger}\}, 
\end{align}
each one of them defined by the block structure  \eqref{blockcs1}, but characterized by different coefficients, $\{f_l,g_l\}$ and $\{\tilde f_l,\tilde g_l\}$. One can easily obtain the Bogoliubov transformation that relates both complex structures:
\begin{align}
\begin{pmatrix} \tilde{a}_{np}^{(x,y)} \\ \tilde{b}^{(x,y)\dagger}_{np} \\ \tilde{a}^{(x,y)\dagger}_{np} \\ \tilde{b}_{np}^{(x,y)}\end{pmatrix}=V_{n}\begin{pmatrix} a_{np}^{(x,y)} \\ b^{(x,y)\dagger}_{np} \\ a^{(x,y)\dagger}_{np} \\ b_{np}^{(x,y)}\end{pmatrix},
\end{align}
with
\begin{align}
V_{n}=\begin{pmatrix} \mathcal{V}_{n} & 0 \\ 0 & \bar{\mathcal{V}}_{n} \end{pmatrix}, \qquad
\mathcal{V}_{n}=\frac{1}{f_{1}^{n}g_{2}^{n}-f_{2}^{n}g_{1}^{n}}\begin{pmatrix}
\tilde{f}_{1}^{n}g_{2}^{n}-\tilde{f}_{2}^{n}g_{1}^{n} & \tilde{f}_{2}^{n}f_{1}^{n}-\tilde{f}_{1}^{n}f_{2}^{n} \\ \tilde{g}_{1}^{n}g_{2}^{n}-\tilde{g}_{2}^{n}g_{1}^{n} & \tilde{g}_{2}f_{1}^{n}-\tilde{g}_{1}f_{2}^{n}
\end{pmatrix}.
\end{align}
Therefore, the beta coefficients of the transformation are given by:
\begin{align}\label{betauniq}
\beta^{h}_{n}(\mathcal{V})=\frac{\tilde{h}_{1}^{n}h_{2}^{n}-\tilde{h}_{2}^{n}h_{1}^{n}}{h_{2}^{n}k_{1}^{n}-h_{1}^{n}k_{2}^{n}},
\end{align}
and, following analogous arguments to those of the previous section, the two complex structures will define unitarily equivalent Fock representations if and only if
\begin{align}
\sum_{n}g_{n}|\beta_{n}^{f}(\mathcal{V})|^{2}<\infty, \qquad \sum_{n}g_{n}|\beta_{n}^{g}(\mathcal{V})|^{2}<\infty.
\end{align}

The denominator of the beta coefficients $\beta^{h}_{n}(\mathcal{V})$ has unit norm owing to the identity \eqref{fgdif}. Moreover, we recall that the considered complex structures have either the coefficient $h_{1}^{n}$ or $h_{2}^{n}$ of order $\omega_{n}^{-1}$, in the ultraviolet regime. We have the following possibilities:
\begin{itemize}
\item[i)] Assume that $h_{1}^{n}$ and $\tilde{h}^{n}_{1}$ are both of order $\omega_{n}^{-1}$. Then, a unitary implementation of the dynamics requires that, in the ultraviolet regime:
\begin{align}
&h_{1}^{n}=\frac{me^{\alpha}}{2\omega_{n}}e^{iH^{n}}+\mathcal{O}(\omega_{n}^{-2}), \qquad h_{2}^{n}=e^{iH^{n}}+\mathcal{O}(\omega_{n}^{-2}),\nonumber\\&
\tilde{h}_{1}^{n}=\frac{me^{\alpha}}{2\omega_{n}}e^{i\tilde{H}^{n}}+\mathcal{O}(\omega_{n}^{-2}), \qquad \tilde{h}_{2}^{n}=e^{i\tilde{H}^{n}}+\mathcal{O}(\omega_{n}^{-2}).
\end{align}
Hence, in this case the beta coefficients are $\beta^{h}_{n}(\mathcal{V})=\mathcal{O}(\omega_{n}^{-2})$, so square summability is guaranteed. From symmetry arguments analogous to those discussed in the previous section, it then follows that $\beta^{k}_{n}(\mathcal{V})$ are square summable as well. Therefore, in the case at hand the two Fock representations coming from the two complex structures $J$ and $\tilde J$ are unitarily equivalent. In complete analogy, unitary equivalence is attained as well when the coefficents of  order $\omega_{n}^{-1}$ are $h_{2}^{n}$ and $\tilde{h}^{n}_{2}$ instead.

\item[ii)] Consider now that $h_{1}^{n}$ and $\tilde{h}^{n}_{2}$ are of order $\omega_{n}^{-1}$. The condition of a unitary quantum dynamics imposes that, in the ultraviolet regime:
\begin{align}
&h_{1}^{n}=\frac{me^{\alpha}}{2\omega_{n}}e^{iH^{n}}+\mathcal{O}(\omega_{n}^{-2}), \qquad h_{2}^{n}=e^{iH^{n}}+\mathcal{O}(\omega_{n}^{-2}),\nonumber\\&
\tilde{h}_{1}^{n}=-e^{i\tilde{H}^{n}}+\mathcal{O}(\omega_{n}^{-2}), \qquad \tilde{h}_{2}^{n}=\frac{me^{\alpha}}{2\omega_{n}}e^{i\tilde{H}^{n}}+\mathcal{O}(\omega_{n}^{-2}).
\end{align}
It then follows that $\beta^{h}_{n}(\mathcal{V})=\mathcal{O}(1)$, and thus the beta coefficients of the Bogoliubov transformation are not square summable. Therefore, in this case the two Fock quantizations coming from the two complex structures would be unitarily inequivalent. Analogously, unitary equivalence is lost as well when the coefficents of  order $\omega_{n}^{-1}$ are $h_{2}^{n}$ and $\tilde{h}^{n}_{1}$ instead. Nevertheless, this inequivalence can be traced back to the fact that we fixed since the beginning the convention for the notions of particle and antiparticle. We can regard these two complex structures as providing opposite conventions. Indeed, the particle-antiparticle interchange $\tilde{a}_{np}^{(x,y)}\longleftrightarrow \tilde{b}_{np}^{(x,y)\dagger}$ implies that the roles of the  coefficients $\tilde{h}^{n}_{l}$ and $\tilde{k}^{n}_{l}$ are interchanged. Since,
according to Eq. \eqref{kcoefs},
\begin{align}
\tilde{k}_{1}^{n}=\frac{me^{\alpha}}{2\omega_{n}}e^{-i\tilde{H}^{n}}e^{i\tilde{G}^{n}}+\mathcal{O}(\omega_{n}^{-2}) \qquad {\rm and} \qquad \tilde{k}_{2}^{n}=e^{-i\tilde{H}^{n}}e^{i\tilde{G}^{n}}+\mathcal{O}(\omega_{n}^{-2})
\end{align}
after this swapping, we are back to case i), and hence unitary equivalence between the redefined complex structure $(\tilde{J})_{h\leftrightarrow k}$ and $J$ is achieved. Actually, the artificial inequivalence between $J$ and $\tilde J$ is easily understandable from a physical point of view, since two theories that treat in an opposite way the concepts of particle and antiparticle will naturally lead to opposite interpretations of physical phenomena.
\end{itemize}

In summary, the invariant complex structures with ultraviolet behavior \eqref{hcoefs} form an unitary equivalence class, up to the convention for the notion of particle and antiparticle. 

In the light of the above discussion, we conclude the following uniqueness result. Within the set of invariant complex structures that present the ultraviolet behavior \eqref{ultrav}, our reference complex structure $J_\text{R}$ is the unique one, up to unitary equivalence, that allows for a unitary implementation of the fermion quantum dynamics. Notice, nonetheless, that in order to reach a fully general result about the uniqueness of the Fock quantization, one still needs to eliminate our restriction on the asymptotics of the complex structures. This will be the subject of future research.

\subsection{Uniqueness of the time-dependent scaling}

Still considering exclusively invariant complex structures, with the block form \eqref{blockcs1}, let us assume that the uniqueness of the Fock representation with unitary implementable dynamics can be proven to be generic, beyond our previous hypothesis on the asymptotics. Namely, let us accept that all invariant Fock representations that allow  for a unitary implementation of the evolution are unitarily equivalent, without any assumption on their ultraviolet asymptotic behavior. In particular, the corresponding equivalence class contains our reference quantization, with complex structure $J_\text{R}$ determined by
\begin{align}
f_{1}^{n}=\frac{me^{\alpha}}{2\omega_{n}}, \qquad f_{2}^{n}=\sqrt{1-(f_1^n)^2}=1+ \mathcal{O}(\omega_n^{-2}), \qquad g_{1}^{n}=f_{2}^{n},\qquad g_{2}^{n}=-f_{1}^{n}.
\end{align}
Any other complex structure $\tilde{J}$ in that equivalence class will be related to $J_\text{R}$ by a Bogoliubov transformation $V_{n}$ with antilinear part of Hilbert-Schmidt class. Hence, the associated beta coefficients, that have the form \eqref{betauniq}, will be square summable over all $n$ and all degeneracies. Let us focus on the case with $h=f$ (a parallel discussion applies for the case with $h=g$ as well). One can easily check that, in the ultraviolet sector,
\begin{align}\label{betaV}
\beta^{h}_{n}(\mathcal{V})=\tilde{h}^{n}_{1}[1+\mathcal{O}(\omega_{n}^{-2})]-e^{i\tilde{H}^{n}}\sqrt{1-|\tilde{h}^{n}_{1}|^{2}}\frac{me^{\alpha}}{2\omega_{n}},
\end{align}
where $\tilde{h}_l^{n}$ are the coefficients associated with the complex structure $\tilde{J}$, and we have used the fact that $\tilde{h}^{n}_{2}$ is fully determined by $\tilde{h}^{n}_{1}$ up to a phase $\tilde H^{n}$, via Eq. \eqref{sympl}. 
By assumption, \mbox{$\sum_n g_n |\beta^{h}_{n}(\mathcal{V})|^2<\infty$}, which implies that  in the ultraviolet regime $\beta^{h}_{n}(\mathcal{V})$ is negligible when compared to $\omega_{n}^{-1}$, since the degeneracy $g_{n}$ is $\mathcal{O}(\omega_{n}^{2})$. Taking into account the restrictions implied by Eq. \eqref{sympl}, one can now distinguish between two possible cases:
\begin{itemize}
\item[i)] Suppose that $\tilde{h}^{n}_{1}$ is of order unity. Then the factor $\exp{(i\tilde{H}^{n})}\sqrt{1-|\tilde{h}^{n}_{1}|^{2}}$ is of the same order as $\tilde{h}^{n}_{1}$, or smaller. It then follows that the dominant contribution in $\beta^{h}_{n}(\mathcal{V})$ comes from the term proportional to $\tilde{h}^{n}_{1}$, which is at least of the order of the unit. But this is a contradiction, inasmuch as unitary equivalence requires $\beta^{h}_{n}(\mathcal{V})$ to be negligible in comparison to $\omega_{n}^{-1}$. We can then rule out this possibility.

\item[ii)] We are left with the case in which $\tilde{h}^{n}_{1}$ is negligible compared to the unit. Here, one gets that
\begin{align}
e^{i\tilde{H}^{n}}\sqrt{1-|\tilde{h}^{n}_{1}|^{2}}\frac{me^{\alpha}}{2\omega_{n}}= e^{i\tilde{H}^{n}}\frac{me^{\alpha}}{2\omega_{n}}+o(\omega_{n}^{-1}),
\end{align}
where $o(\omega_{n}^{-1})$ stands for terms negligible with respect to $\omega_{n}^{-1}$. Hence, since $\beta^{h}_{n}(\mathcal{V})$ is also negligible compared to $\omega_{n}^{-1}$, we must have
\begin{align}\label{scaling}
\tilde{h}^{n}_{1}=  e^{i\tilde{H}^{n}}\frac{me^{\alpha}}{2\omega_{n}}+o(\omega_{n}^{-1}).
\end{align}
Namely, the dominant term in $\tilde{h}^{n}_{1}$ is fixed, up to the phase $\tilde H^{n}$. This fixation is immediately inherited, up to phases, by the rest of coefficients that determine the complex structure $\tilde{J}$, via the relations \eqref{hcoefs}-\eqref{kcoefs}.
\end{itemize}

As a consequence of this analysis, we conclude  the following. Under the assumption that all invariant Fock representations that allow  a unitary implementation of the dynamics are unitarily equivalent, it follows that the time-dependent scalings of the dominant terms (in the asymptotic limit of large $\omega_{n}$) in the particle and antiparticle parts of the fermion field are essentially unique, up to phases.

Furthermore, we have just proven that if any invariant Fock quantization were to be unitarily equivalent to our reference one, then its asymptotic behavior must be exactly the same one as that of our reference quantization, namely, the one given by Eq. \eqref{hcoefs}, at least up to terms $o(\omega_{n}^{-1})$ in $h_1^n$ (and its $k$-counterpart). In other words, if an invariant complex structure $\tilde J$ displays a different asymptotic behavior than that of $J_\text{R}$ (at the considered orders), then the quantum theories that $\tilde J$ and $J_\text{R}$ define are necessarily inequivalent.

A final remark is in order at this point. Let us recall that all the Fock quantizations analyzed so far are those that concern the fermion spinor scaled by a time-dependent factor $\exp({3\alpha/2})$, according to the expansions \eqref{harm1}-\eqref{harm4}. Let us denote the non-scaled modes as $\tilde{x}_{np}=x_{np}\exp({-3\alpha/2})$ and $\tilde{y}_{np}=y_{np}\exp({-3\alpha/2})$. Notice that the non-zero Dirac brackets of such modes are given by
\begin{align}
\{\tilde{x}_{np},\bar{\tilde{x}}_{np}\}=-ie^{-3\alpha},\qquad \{\tilde{y}_{np},\bar{\tilde{y}}_{np}\}=-ie^{-3\alpha}.
\end{align}
Imagine now that we were to consider the invariant Fock representations of the original (non-scaled) fermion field, namely, the choices of annihilation and creation-like variables:
\begin{align}\label{blockcs2}
\begin{pmatrix} a_{np}^{(\tilde{x},\tilde{y})} \\ b^{(\tilde{x},\tilde{y})\dagger}_{np} \\ a^{(\tilde{x},\tilde{y})\dagger}_{np} \\ b_{np}^{(\tilde{x},\tilde{y})}\end{pmatrix}_{\!\!\eta}=\begin{pmatrix}\tilde{f}_{1}^ {n}(\eta) & \tilde{f}_{2}^{n}(\eta) & 0 & 0 \\ \tilde{g}_{1}^ {n}(\eta) & \tilde{g}_{2}^{n}(\eta) & 0 & 0 \\ 0 & 0 & \bar{\tilde{f}}^{n}_{1}(\eta) & \bar{\tilde{f}}^{n}_{2}(\eta) \\ 0 & 0 & \bar{\tilde{g}}^{n}_{1}(\eta) & \bar{\tilde{g}}^{n}_{2}(\eta)\end{pmatrix}\begin{pmatrix} \tilde{x}_{np} \\ \bar{\tilde{y}}_{np} \\ \bar{\tilde{x}}_{np} \\ \tilde{y}_{np}\end{pmatrix}_{\!\!\eta},
\end{align}
with the conditions
\begin{align}\label{sympl2}
|\tilde{f}_{1}^{n}|^{2}+|\tilde{f}_{2}^{n}|^{2}=e^{3\alpha}, \qquad |\tilde{g}_{1}^{n}|^{2}+|\tilde{g}_{2}^{n}|^{2}=e^{3\alpha}, \qquad \tilde{f}_{1}^{n}\bar{\tilde{g}}^{n}_{1}+\tilde{f}_{2}^{n}\bar{\tilde{g}}^{n}_{2}=0,
\end{align}
for these variables to be indeed annihilation and creation-like. One may always write the coefficients as $\tilde{f}_{l}^{n}=f_{l}^{n}\exp({3\alpha/2})$ and $\tilde{g}_{l}^{n}=g_{l}^{n}\exp({3\alpha/2})$,
where $f_{l}^{n}$ and $g_{l}^{n}$ satisfy the relations \eqref{sympl}. From our analysis above, if we assume uniqueness of the quantization when restricted to invariant complex structures that allow for a unitary implementation of the dynamics, then the time-dependent scaling is fixed in the dominant terms (with respect to the asymptotics of large $\omega_{n}$) of $f_{l}^{n}$ and $g_{l}^{n}$, and hence so is for their counterparts $\tilde{f}_{l}^{n}$ and $\tilde{g}_{l}^{n}$. Notice that such a choice of annihilation and creation-like variables is equivalent to a quantization of the original field in Eqs. \eqref{harm1}-\eqref{harm4}, since the scaling factor is global. Therefore, we can conclude here that, if there is a general uniqueness result for the invariant Fock representations that admit a unitary dynamics, this result implies that the global scaling of the field by $\exp{(3\alpha/2)}$ is necessary.

\section{Conclusions and outlook}
\label{sec:conclu}

In this work, we have dealt with part of the infinite ambiguity that is always present in the quantum description of any field-like system, and in particular in the case of a Dirac fermion field propagating in a cosmological spacetime. Even in the case of a Fock quantization for fields with linear dynamical equations, there are generically two sources for this ambiguity. First of all, the choice of the variables parameterizing  the field is always subject to the freedom of including any part of the field evolution into the time dependence of the considered spacetime, via a time-dependent scaling of either the whole or just a part of the field. On the other hand, even when particular configuration variables have been selected, there is an infinite freedom in the choice of a complex structure, and therefore, in the choice of the one-particle Hilbert space of the quantum theory. This generically leads to an infinite number of unitarily inequivalent quantum theories, that therefore describe infinitely many different possible physics. 

More specifically, we have investigated the Fock quantization of a purely inhomogeneous massive Dirac field coupled to an expanding FRW cosmology with $S^{3}$ topology of the spatial sections, model which had been previously studied in Ref. \cite{H-D}. We have shown that the combined criteria of invariance of the representation under the symmetries of the field equations, and of unitary implementability of the fermion dynamics, restrict the set of allowed vacua, even to the point of achieving a unique class of unitarily equivalent quantum theories under some further mild condition on the considered quantizations. This conditon requires that the complex structure possesses an asymptotic behavior in the ultraviolet limit of large Dirac eigenvalues, characterized by the first identity in Eq. \eqref{ultrav} [the second formula in that equation and Eq. \eqref{ultravk} follow from the former via relations \eqref{sympl}-\eqref{fgrel}]. 

Within this set of complex structures, we have derived the necessary and sufficient condition that they need to satisfy to allow for a unitary implementation of the dynamics in the corresponding Fock representation. The result is that the asymptotic limit must verify condition \eqref{unitcond2}. This in turn means that
the time-dependent factors of the dominant terms (in the limit of large Dirac eigenvalues) of the particle and antiparticle parts of the field, and for the two chiralities, are uniquely fixed (up to phases). 

It is worth emphasizing that the Fock quantization selected in Ref. \cite{H-D} belongs to this equivalence class that allows for a unitary implementation of the dynamics, property that underlies the finite production of particles and antiparticles on the evolved vacuum that was found in that work. The choice of quantization made in Ref. \cite{H-D} has motivated us to introduce a reference complex structure, that can be considered the simplest one belonging to the constructed equivalence class.

We have seen that the unitary equivalence within this class is found only after adapting the convention of what is a particle and what is an antiparticle. This is just a manifestation of the fact that two theories with an opposite notion for particles and antiparticles will differ in the physical interpretation of the results if their conventions are not reconciled. 

Our uniqueness result for the quantization, though rather general, is not still fully so, because the considered set of Fock representations is not the most general one satisfying the physical criteria of invariance of the vacuum under the symmetries of the dynamical equations and of a unitary evolution, inasmuch as we have assumed certain type of asymptotic behavior in the ultraviolet sector. Nonetheless, our results point out to the possibility that this assumption may in fact not be too restrictive once our criteria are imposed. Indeed, we have shown that any other Fock quantization selected by those criteria and which is unitarily equivalent to our reference one must display an ultraviolet behavior similar to that assumed here. Therefore, a possible line of attack to complete the proof of uniqueness would be to discuss the extent to which unitary implementability of the dynamics forbids other types of ultraviolet behaviors. We will address this issue somewhere else.

This uniqueness result provides a strong physical guideline for the selection of a Fock quantization of a fermion Dirac field in an FRW cosmology with closed spatial sections. Furthermore, its extension to more realistic spatial topologies, such as flat ones, would be extremely useful in the analysis of the possible consequences of the presence of fermion matter perturbations in the early stages of the universe. In particular, this is relevant for the analysis of the cosmic neutrino background. Note that, in order to investigate the extension of our results to other topologies for the spatial sections, we just need to consider the Dirac operator for those sections, with its associated mode decomposition and symmetries. An asymptotic analysis similar to the one carried out here would elucidate whether uniqueness is as well guaranteed by our criteria with these other spatial topologies.

The uniqueness of the Fock quantization is also of great advantage when going beyond the framework of quantum field theory in curved cosmological spacetimes, to that of quantum cosmology, as in Ref. \cite{H-D}. Indeed, in that work the homogeneous modes that provide the cosmological background are quantized adopting a standard Schr\"odinger representation, and the question of backreaction between this background and the fermionic perturbations is investigated. It would be interesting to perform a similar analysis adopting for the zero modes of the geometry the representation of Loop Quantum Cosmology \cite{lqc}, as it has been already done for the case of scalar perturbations  \cite{inf-hybrid,inf-ash}. This would extend the analysis of fermionic perturbations in cosmology beyond the onset of inflation.

Additionally, we could also try to extend our results to account for other types of fermion fields (such as Weyl fermions, massless Dirac fermions, or Majorana fermions), as well as other background geometries. These studies would be particularly relevant in the context of condensed matter physics, where the experimental use of fermion excitations may give rise to a breakthrough in areas such as electronics and quantum computing. For instance, in the physics of graphene, it is known that the low energy electronic excitations can be described by a massless two dimensional Dirac equation \cite{graphene}. Other examples are the increasing evidence of the existence of Majorana fermions in certain superconductors coupled to nanowires \cite{majorana}, and of Weyl fermion semimetals \cite{weyl}. 

\acknowledgments
This work was partially supported by the research grants MICINN/MINECO Project No. FIS2011-30145-C03-02 and its continuation FIS2014-54800-C2-2-P from Spain, DGAPA-UNAM IN113115 and CONACyT 237351 from Mexico, and COST Action MP1405 QSPACE, supported by COST (European Cooperation in Science and Technology). In addition, M. M-B acknowledges financial support from the Netherlands Organisation for Scientific Research (NWO) (Project No. 62001772).

\appendix

\section{Weyl representation of the Dirac matrices}
\label{appendixA1}

Here we provide the specific form of the Weyl representation taken for the Dirac matrices in this work. Specifically, the convention for their definition as a representation of the Dirac-Clifford algebra is the following:
\begin{align}
\gamma^{a}\gamma^{b}+\gamma^{b}\gamma^{a}=2\eta^{ab}I,
\end{align}
where $I$ is the $4\times 4$ identity matrix and $\eta^{ab}$ is the Minkowski metric, given in Cartesian coordinates by $\text{diag}\{-1,+1,+1,+1\}$. In the Weyl representation, these matrices take the explicit expression  
\begin{align}
\gamma^{a}=i\begin{pmatrix}
0 & \Sigma^{a} \\ \widetilde{\Sigma}^{a} & 0
\end{pmatrix},
\end{align}
where $\Sigma^{0}=\widetilde{\Sigma}^{0}$ is the $2\times 2$ identity matrix, and $\Sigma^{j}=-\widetilde{\Sigma}^{j}$ with $j=1,2,3$ are the standard Pauli matrices, namely
\begin{align}
\Sigma^{1}=\begin{pmatrix}
0 & 1 \\ 1& 0
\end{pmatrix},\qquad
\Sigma^{2}=\begin{pmatrix}
0 & -i \\ i& 0
\end{pmatrix},\qquad
\Sigma^{3}=\begin{pmatrix}
1 & 0 \\ 0& -1
\end{pmatrix}.
\end{align}

\section{Aymptotics of the dynamics}
\label{appendixA}

The dynamics of the modes $\{z_{np}\}=\{x_{np},y_{np}\}$ is governed by the second-order equation
\begin{align}\label{app2order}
z_{np}''=\alpha 'z_{np}'-(\omega^2_{n}+m^{2}e^{2\alpha}+i\omega_{n}\alpha ')z_{np}.
\end{align}
Let us call $\tilde{z}_{np}=z_{np}\exp({-\Delta{\alpha}/2})$, with $\Delta{\alpha}=\alpha-\alpha_{0}$. In terms of this scaled mode, the equation of motion reads
\begin{align}
\tilde{z}_{np}''+\left(\tilde{\omega}_{n}^{2}+m^{2}e^{2\alpha}+\frac{\alpha''}{2}\right)\tilde{z}_{np}=0,
\end{align}
where $\tilde{\omega}_{n}=\omega_{n}+i\alpha'/2$. Let us search for two independent solutions $\tilde{z}_{np}^{l}$, with $l=1,2$, of the form
\begin{align}\label{applaurent}
\tilde{z}_{np}^{l}=\exp{\left[-i(-1)^{l}\widetilde{\Theta}_{n}^{l}\right]}, \qquad \text{with} \qquad (\widetilde{\Theta}_{n}^{l})'=\tilde{\omega}_{n}+v_{n}^{l}+\Lambda_{n}^{l},
\end{align}
and where $v^{l}_{n}=\mathcal{O}(1)$ and $\Lambda^{l}_{n}=\mathcal{O}(\omega_{n}^{-1})$ in the ultraviolet limit of large $\omega_{n}$. Introducing this ansatz in the equation of motion one obtains
\begin{align}
0=&i(-1)^{l}[(\Lambda^{l}_{n})'+(v_{n}^{l})']+2\tilde{\omega}_{n}v_{n}^{l}
+2\tilde{\omega}_{n}\Lambda^{l}_{n}+(v^{l}_{n})^{2}\nonumber\\&+(\Lambda^{l}_{n})^{2}+2v_{n}^{l}\Lambda_{n}^{l}
-[(-1)^{l}+1]\frac{\alpha ''}{2}-m^{2}e^{2\alpha}.
\end{align} 
In the ultraviolet sector, at highest order in $\omega_{n}$, one then finds
\begin{align}
2\tilde{\omega}_{n}v_{n}^{l}=0.
\end{align}
Therefore, $v^{l}_{n}$ must be zero (recall that it has been assumed to be of order unity). Taking this result into account, the remaining equation of motion turns out to be a first-order differential equation for $\Lambda_{n}^{l}$. Letting $\Lambda_{n}^{l}(\eta_{0})=0$, the solution to that equation can indeed be seen to be of order $\omega_{n}^{-1}$ (see Appendix \ref{appendixB}). Hence, the equation of motion for the scaled modes admits two independent solutions of the form \eqref{applaurent} with initial conditions for the derivative of $\widetilde{\Theta}^{l}_{n}$ equal to $\tilde{\omega}_n$.

Let us recall that we had called $\exp({i\Theta^{1}_{n}})$ and $\exp({-i\Theta^{2}_{n}})$ the two independent solutions of the equation of motion \eqref{app2order} for the original modes. Thus, inserting the solutions \eqref{applaurent} for the scaled modes and relating them with the non-scaled ones, we get
\begin{align}
(\Theta^{l}_{n})'=(\widetilde{\Theta}^{l}_{n})'+\frac{i}{2}(-1)^{l}\alpha '=\omega_{n}+\frac{i}{2}[1+(-1)^{l}]\alpha '+\Lambda^{l}_{n}.
\end{align}
Therefore, with the initial conditions $\Theta_{n,0}^{l}=\Theta_{n}^{l}(\eta_{0})=0$, the phases of the solutions turn out to be
\begin{align}
\Theta^{l}_{n}=\omega_{n}\Delta\eta+\frac{i}{2}[1+(-1)^{l}]\Delta\alpha +\int_{\eta_{0}}^{\eta}d\tilde\eta\,\Lambda^{l}_{n}(\tilde\eta),
\end{align}
with $\Delta\eta=\eta-\eta_{0}$. The initial conditions for their derivatives, inherited from the ones for the scaled solutions, are given by:
\begin{align}
\Theta^{l}_{n,1}=(\widetilde{\Theta}^{l}_{n})'(\eta_{0})+\frac{i}{2}(-1)^{l}\alpha_{0}'=\omega_{n}+\frac{i}{2}[1+(-1)^{l}]\alpha_{0}'.
\end{align}

These initial conditions in turn translate into the following values for the quantities introduced in Eq. \eqref{icconsts}:
\begin{align}
\Delta^{1}_{n}=0,\qquad \Delta^{2}_{n}=1,\qquad \Gamma^{l}_{n}=\frac{me^{\alpha_{0}}}{2\omega_{n}+i\alpha_{0}'}=\frac{me^{\alpha_{0}}}{2\omega_{n}}+\mathcal{O}(\omega_{n}^{-2}),
\end{align}
where the last equality holds for large $\omega_{n}$.

\section{Asymptotic behavior of $\Lambda_{n}^{l}$}
\label{appendixB}

The phases $\Lambda_{n}^{l}$ obey the first-order differential equation of Riccati type
\begin{align}\label{lambdaeq}
(\Lambda_{n}^{l})'=i(-1)^{l}[(\Lambda_{n}^{l})^{2}+(2{\omega}_{n} +i\alpha')\Lambda_{n}^{l}]-u^{l},
\end{align}
where we have defined the following time-dependent and $\omega_{n}$-independent functions:
\begin{align}
u^{l}(\eta)=i[(-1)^{l}+1]\frac{\alpha''(\eta)}{2}+i(-1)^{l}m^{2}e^{2\alpha(\eta)}.
\end{align}

We proceed to analyze the asymptotics of this equation, in the limit of large $\omega_{n}$. In this limit, we can start by neglecting the quadratic term on $\Lambda_{n}^{l}$, since it would be dominated by the linear one. Let us then study the solutions $\widetilde{\Lambda}_{n}^{l}$ of
\begin{align}
(\widetilde{\Lambda}_{n}^{l})'=(-1)^{l}(2i\omega_{n}-\alpha ')\widetilde{\Lambda}_{n}^{l}-u^{l}.
\end{align}
The solutions with initial condition $\widetilde{\Lambda}_{n}^{l}(\eta_{0})=0$ are
\begin{align}
\widetilde{\Lambda}_{n}^{l}=-\exp[(-1)^{l}(2i\omega_{n}\eta-\alpha)]\int_{\eta_{0}}^{\eta} d\tilde{\eta}\,u^{l}(\tilde{\eta})\exp\{(-1)^{l}[\alpha(\tilde{\eta})-2i\omega_{n}\tilde{\eta}]\}.
\end{align}
Integration by parts yields
\begin{align}
\widetilde{\Lambda}_{n}^{l}=&\frac{i(-1)^{l}}{2\omega_{n}}\bigg\{u^{l}_{0}\exp[(-1)^{l}(2i\omega_{n}\Delta\eta-\Delta\alpha)]-u^{l}+\exp[(-1)^{l}(2i\omega_{n}\eta-\alpha)]\nonumber\\&\times\int_{\eta_{0}}^{\eta}d\tilde{\eta}\,[(u^{l})'(\tilde{\eta})+(-1)^{l}u^{l}(\tilde{\eta})\alpha '(\tilde{\eta})]\exp\{(-1)^{l}[\alpha(\tilde{\eta})-2i\omega_{n}\tilde{\eta}]\}\bigg\},
\end{align}
with $u^{l}_{0}=u^{l}(\eta_{0})$. Therefore, since neither $\alpha$ nor $u^{l}$ depend on $\omega_{n}$, if both $(u^{l})'$ and $u^{l}\alpha '$ exist and are integrable in every closed interval $[\eta_{0},\eta]$, then there exists a positive function $C(\eta)$ which is $\omega_{n}$-independent and such that the absolute value of $\widetilde{\Lambda}_{n}^{l}(\eta)$ is bounded by $C(\eta)/\omega_{n}$.

Now, let us notice that $(\widetilde{\Lambda}_{n}^{l})^{2}(\eta)$ is bounded by $C(\eta)^{2}/\omega_{n}^{2}$, and it is hence negligible in the ultraviolet limit when compared to the linear term in Eq. \eqref{lambdaeq}. We thus conclude that the functions $\widetilde{\Lambda}_{n}^{l}=\mathcal{O}(\omega_{n}^{-1})$ can be taken as asymptotic solutions of \eqref{lambdaeq} in the limit of large $\omega_{n}$, up to higher-order corrections.

\end{document}